\documentclass[prd,preprint,showpacs,preprintnumbers,nofootinbib,eqsecnum,superscriptaddress]{revtex4}

 \usepackage[dvips,final]{graphicx}
  \usepackage{amssymb}
   \usepackage{amsmath}
    \usepackage{amsfonts}
     \usepackage{epsfig}
      \usepackage{bm}

\usepackage{mathpazo}

\usepackage[section]{placeins}

\usepackage{multirow}
\usepackage{ctable}
\usepackage{booktabs}
\usepackage{array}
\usepackage{tabularx}
\usepackage{xcolor}
\usepackage{pstricks}

\begin{document}

\title{Prompt inclusive production of $J/\psi$, $\psi'$ and $\chi_{c}$ mesons
at $\sqrt{s}$ $=$ $68.5$ $GeV$ energy within NRQCD $k_t$-factorization approach}

\author{Anna Cisek}
\email{acisek@ur.edu.pl}
\affiliation{Faculty of Exact and Technical Sciences,
University of Rzesz\'ow, ul. Pigonia 1, PL-35-310 Rzesz\'ow, Poland}

\author{Antoni Szczurek}
\email{antoni.szczurek@ifj.edu.pl} 
\affiliation{Institute of Nuclear Physics, Polish Academy of Sciences, 
ul. Radzikowskiego 152, PL-31-342 Krak{\'o}w, Poland}

\begin{abstract}
We discuss prompt production of $J/\psi$ mesons in proton-proton
collisions at the $\sqrt{s}$ $=$ $68.5$ $GeV$ energy within NRQCD
$k_t$-factorization approach
using different unintegrated gluon distributions functions (UGDFs).
We include both direct color-singlet production ($g g \to J/\psi g$) 
as well as a feed-down from $\chi_c \to J/\psi \gamma$ and 
$\psi' \to J/\psi X$ decays.
The production of the decaying mesons $\chi_c$ or $\psi'$ is also
calculated within NRQCD $k_t$-factorization approach.
Differential distributions in rapidity and transverse momentum of $J/\psi$
are calculated and compared with experimental data of the LHCb 
collaborations. The LHCb fixed target data allow to test UGDFs
in a new range of $x$.
\end{abstract}

\pacs{12.38.-t, 13.60Le, 13.85Ni, 14.40.-n}
\maketitle

\section{Introduction}

The production of $J/\psi$ in proton-proton collisions is known as 
a rather chalanging task. The color singlet leading-order collinear 
approach gives much smaller cross section than measured at Fermilab 
or the LHC. Often color octet contributions must be introduced to describe
experimental data. Different authors use different long distance
matrix elements to get a satisfactory description of the data.
Here we shall concentrate on prompt $J/\psi$ production, i.e. the
decays of $b$ and $\bar b$ (or rather $B$ and $\bar B$) will be neglected.
The direct production is not sufficient and one has to include also production
and decays of other quarkonia which give sizeable contribution.

Some tim ago we have shown that the $k_t$-factorization approach
with nonrelativistic approximation and unintegrated gluon distributions
provide a quite good description of the world data \cite{CS2018}.
This is because the $k_t$-factorization approach includes effectively
higher-order QCD corrections.

In the present studies we wish to test how good is such an approach
at lower energies. A few years ago the LHCb collaboration presented
a first result for so-called fixed target experiments using the so-called SMOG device.
In this analysis neon target was used and the data was devided by $A=$20
and present for one nucleon.
An updated result was published in \cite{LHCb_jpsi} (see also \cite{Mattioli}).
So far the unintegrated UGDFs were tested in different processes
rather at energies in which one is sensitive to the region
of not too high longitudinal momentum fractions carried by gluon
in the proton (nucleon). The region of UGDFs at larger values of $x$
was not well tested so far.
Therefore the relatively new fixed target data give a chance for such
tests. A first test will be done here.

\section{Formalism}

In our calculation we completely neglect all nuclear effects
that should be rather small for $A=$20.

\subsection{Direct production}

\begin{figure}
\includegraphics[width=6.5cm]{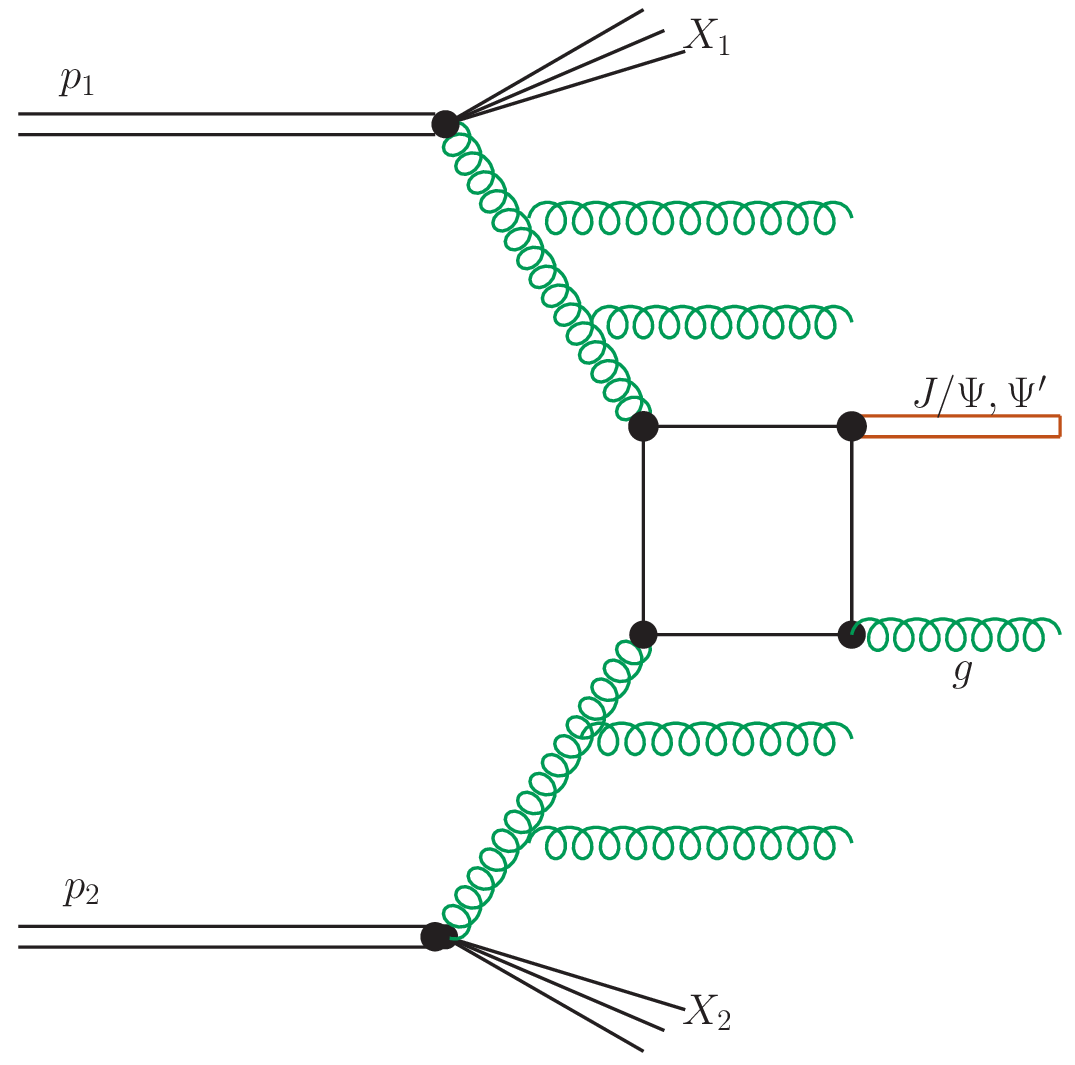}
\caption{The diagram for direct $J/\psi$ ($\psi'$) meson production
in the $k_t$-factorization approach.}
\label{fig:gg_Jpsig}
\end{figure}

The main color-singlet mechanism of $J/\psi$ meson production is 
illustrated in Fig.\ref{fig:gg_Jpsig}. In this case $J/\psi$ is 
produced directly in association with an extra ``hard'' gluon due 
to C-parity conservation.

We calculate the dominant color-singlet $g g \to J/\psi g$
contribution taking into account transverse momenta of initial gluons.
In the $k_t$-factorization approach the differential cross section can 
be written as:
\begin{eqnarray}
\frac{d \sigma(p p \to J/\psi g X)}{d y_{J/\psi} d y_g d^2 p_{J/\psi,t} d^2 p_{g,t}}
&& = 
\frac{1}{16 \pi^2 {\hat s}^2} \int \frac{d^2 q_{1t}}{\pi} \frac{d^2 q_{2t}}{\pi} 
\overline{|{\cal M}_{g^{*} g^{*} \rightarrow J/\psi g}^{off-shell}|^2} 
\nonumber \\
&& \times \;\; 
\delta^2 \left( \vec{q}_{1t} + \vec{q}_{2t} - \vec{p}_{H,t} - \vec{p}_{g,t} \right)
{\cal F}_g(x_1,q_{1t}^2,\mu^2) {\cal F}_g(x_2,q_{2t}^2,\mu^2)  ;
\label{kt_fact_gg_jpsig}
\end{eqnarray}
where ${\cal F}_g$ are unintegrated gluon distributions functions.

The corresponding matrix element squared for the $g g \to J/\psi g$ is
\begin{equation}
|{\cal M}_{gg \to J/\psi g}|^2 \propto \alpha_s^3 |R(0)|^2 \; .
\label{matrix_element} 
\end{equation}
Running coupling constants are used in the present calculation. 
Different combinations of renormalization scales were tried. 
We decided to use:
\begin{equation}
\alpha_s^3 \to \alpha_s(\mu_1^2) \alpha_s(\mu_2^2) \alpha_s(\mu_3^2) \; ,
\end{equation}
where $\mu_1^2 = max(q_{1t}^2,m_t^2)$,
      $\mu_2^2 = max(q_{2t}^2,m_t^2)$ and
      $\mu_3^2 = m_t^2$.
Here $m_t$ is the $J/\psi$ transverse mass.
The factorization scale in the calculation was taken as
$\mu_F^2 = (m_t^2 + p_{t,g}^2)/2$.

\subsection{$J/\psi$ from decays}

\begin{figure}
\begin{center}
\includegraphics[width=6.5cm]{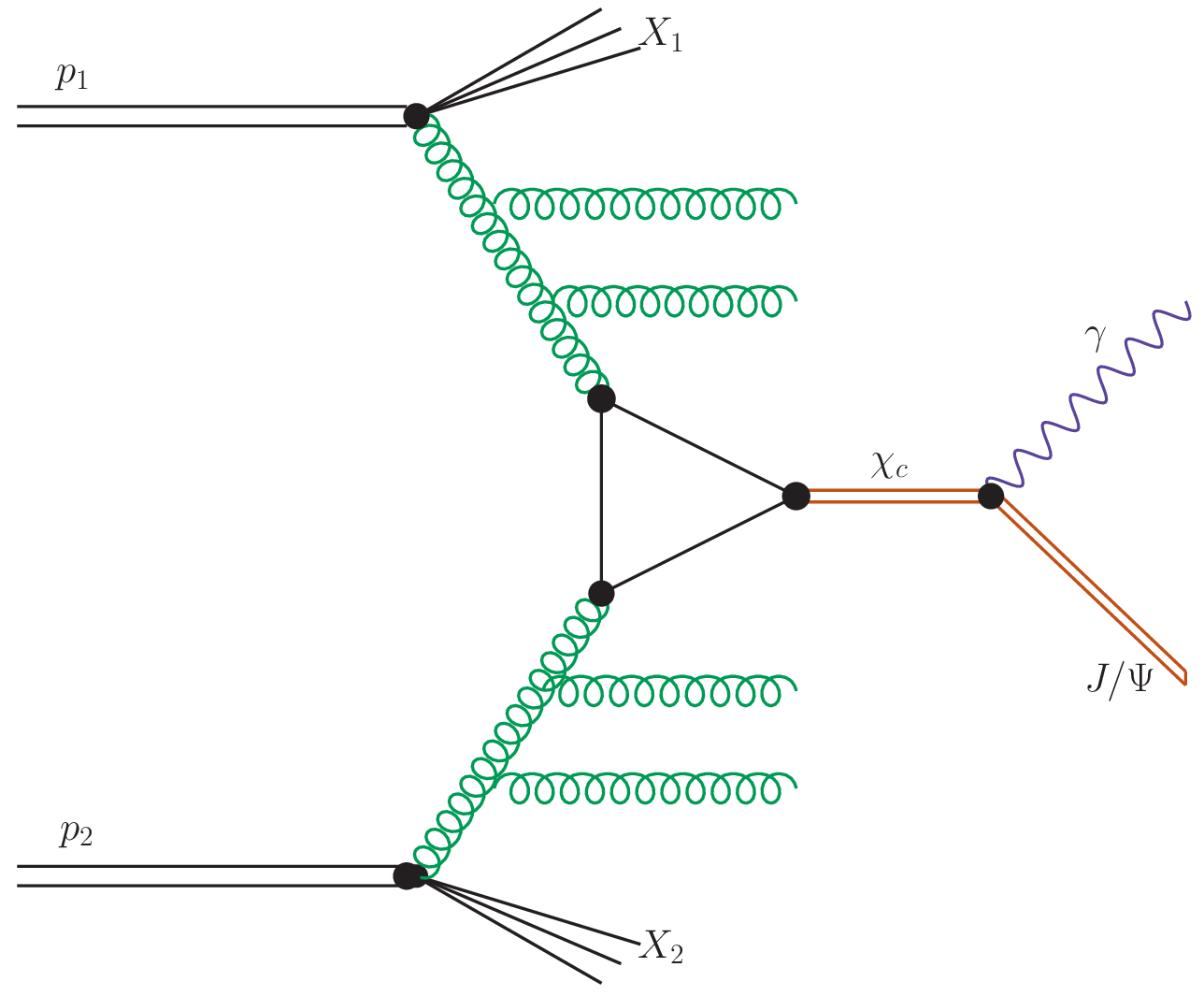}
\end{center}
\caption{The leading-order diagram for $\chi_c$ meson production
and decay in the $k_t$-factorization approach.}
\label{fig:gg_chic}
\end{figure}

Similarly we do calculation for P-wave $\chi_c$ meson production.
Here the lowest-order subprocess $g g \to \chi_c$ is allowed by
positive $C$-parity of $\chi_c$ mesons.
In the $k_t$-factorization approach the leading-order cross section 
for the $\chi_c$ meson production can be written somewhat formally as:
\begin{eqnarray}
\sigma_{pp \to \chi_c} = \int \frac{dx_1}{x_1} \frac{dx_2}{x_2}
\frac{d^2 q_{1t}}{\pi} \frac{d^2 q_{2t}}{\pi} 
&&\delta \left( (q_1 + q_2)^2 - M_{\chi_c}^2 \right) 
\sigma_{gg \to H}(x_1,x_2,q_{1},q_{2}) \nonumber \\
&&\times \; {\cal F}_g(x_1,q_{1t}^2,\mu_F^2) {\cal F}_g(x_2,q_{2t}^2,\mu_F^2)
\; ,
\label{chic_kt_factorization}
\end{eqnarray}
where ${\cal F}_g$ are unintegrated (or transverse-momentum-dependent) 
gluon distributions and $\sigma_{g g \to \chi_c}$ is 
$g g \to \chi_c$ (off-shell) cross section.
The situation is illustrated diagramatically in Fig.\ref{fig:gg_chic}.

The matrix element squared for the $g g \to \chi_c$ subprocess is
\begin{equation}
|{\cal M}_{gg \to \chi_c}|^2 \propto \alpha_s^2 |R'(0)|^2 \; .
\label{matrix_element} 
\end{equation}

After some manipulation: 
\begin{eqnarray}
\sigma_{pp \to \chi_c} = \int d y d^2 p_t d^2 q_t \frac{1}{s x_1 x_2}
\frac{1}{m_{t,\chi_c}^2}
\overline{|{\cal M}_{g^*g^* \to \chi_c}|^2} 
{\cal F}_g(x_1,q_{1t}^2,\mu_F^2) {\cal F}_g(x_2,q_{2t}^2,\mu_F^2) / 4
\; ,
\label{useful_formula}
\end{eqnarray}
that can be also used to calculate rapidity and transverse
momentum distribution of the $\chi_c$ mesons.

In the last equation:
$\vec{p}_t = \vec{q}_{1t} + \vec{q}_{2t}$ is transverse momentum 
of the $\chi_c$ meson
and $\vec{q}_t = \vec{q}_{1t} - \vec{q}_{2t}$ is an auxiliary variable 
which is used for the integration of the cross section. Furthermore:
$m_{t,{\chi_c}}$ is the so-called $\chi_c$ transverse mass and
$x_1 = \frac{m_{t,\chi_c}}{\sqrt{s}} \exp( y)$,  
$x_2 = \frac{m_{t,\chi_c}}{\sqrt{s}} \exp(-y)$.
The factor $\frac{1}{4}$ is the Jacobian of transformation from
$(\vec{q}_{1t}, \vec{q}_{2t})$ to $(\vec{p}_t, \vec{q}_{t})$ variables.

As for the $J/\psi$ production running coupling contants are used. 
Different combination of scales were tried. The best choices are:
\begin{equation}
\alpha_s^2 \to \alpha_s(\mu_1^2) \alpha_s(\mu_2^2) \; ,
\end{equation}
where 
      $\mu_1^2 = max(q_{1t}^2,m_t^2)$ and
      $\mu_2^2 = max(q_{2t}^2,m_t^2)$.
Above $m_t$ is transverse mass of the $\chi_c$ meson.
The factorization scale(s) for the $\chi_c$ meson production are 
fixed traditionally as $\mu_F^2 = m_t^2$.

The $J/\psi$ mesons are produced then by the $\chi_c \to J/\psi \gamma$
decays which are dominated by the E1 transitions.

\section{Results}

In this presentation we will do calculations with the following
unintegrated gluon distribution functions used previously in
the literature:

\begin{itemize} 
\item (a) Kimber-Martin-Ryskin (KMR) (see \cite{KMR2001,HMMT2015}),
\item (b) Jung-Hautmann (JH2013) (see \cite{JH2013}), 
\item (c) Gaussian with $\sigma = 0.5$ GeV and CTEQ-Tea Parton 
          Distribution Functions (see \cite{CTEQ-Tea}),
\item (d) Kharzeev-Levin (KL) (see \cite{KL}), 
\item (e) Kutak-Stasto (KS) (see \cite{KS}), 
\item (f) Moriggi-Peccini-Machado (MPM) (see \cite{MPM}).
\end{itemize}

Some of those UGDFs give good description of different data
for larger energies, i.e. smaller values of longitudinal
momentum fractions. The KMR distribution gives good description
of different data in a broad range of energies. The JH2013 UGDF
was tested for inclusive production of $D$ mesons for p+A
fixed target collisions (see \cite{MS2022a,MS2022b,GMS2024}).

Let us start from analysing $x_1$, $x_2$ values relevant for
production of $\chi_c(1^+)$, $\chi_c(2^+)$ and $J/\psi$ production.

\begin{figure}[h]
\centering
\includegraphics[width=7.cm]{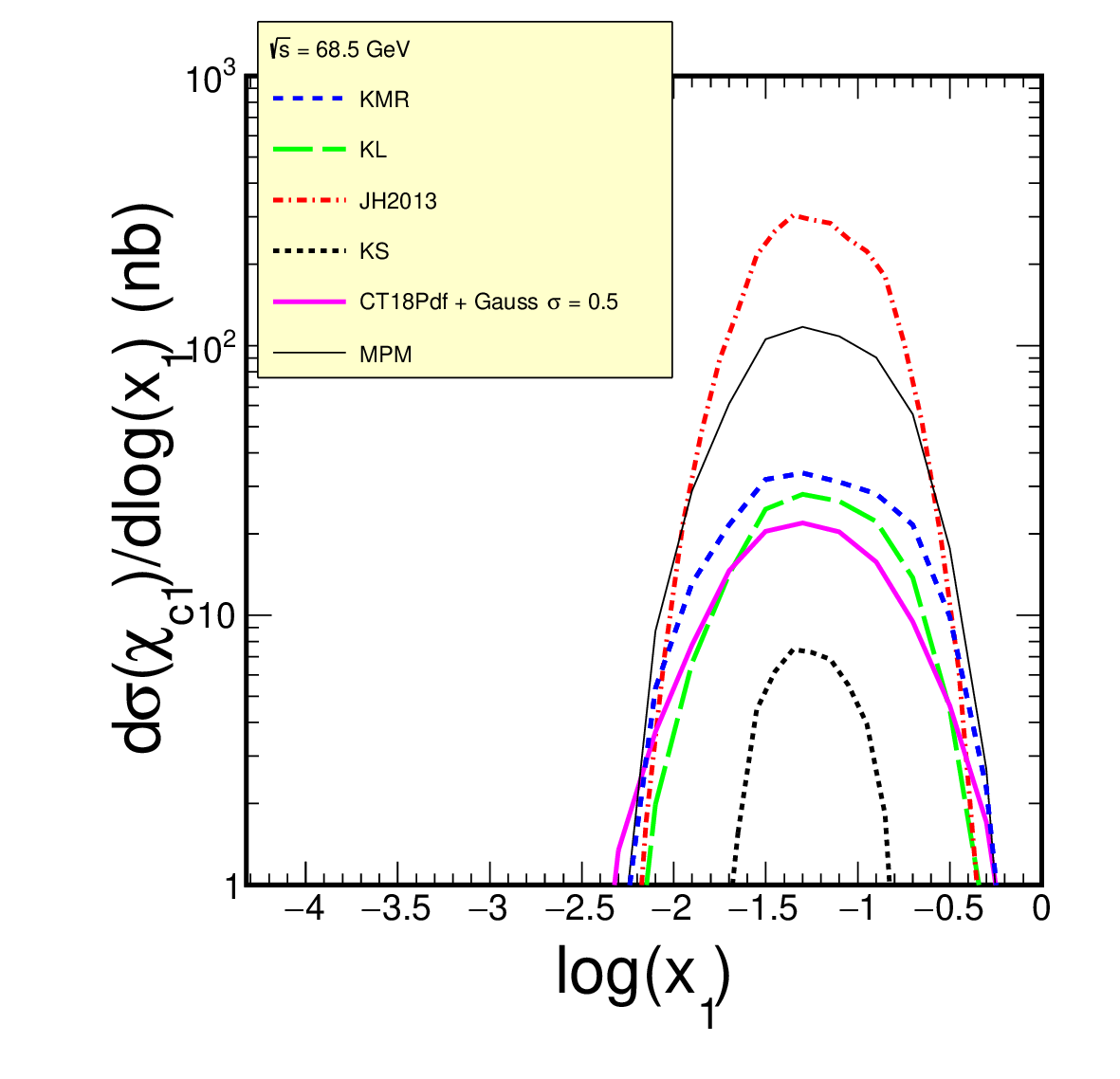}
\includegraphics[width=7.cm]{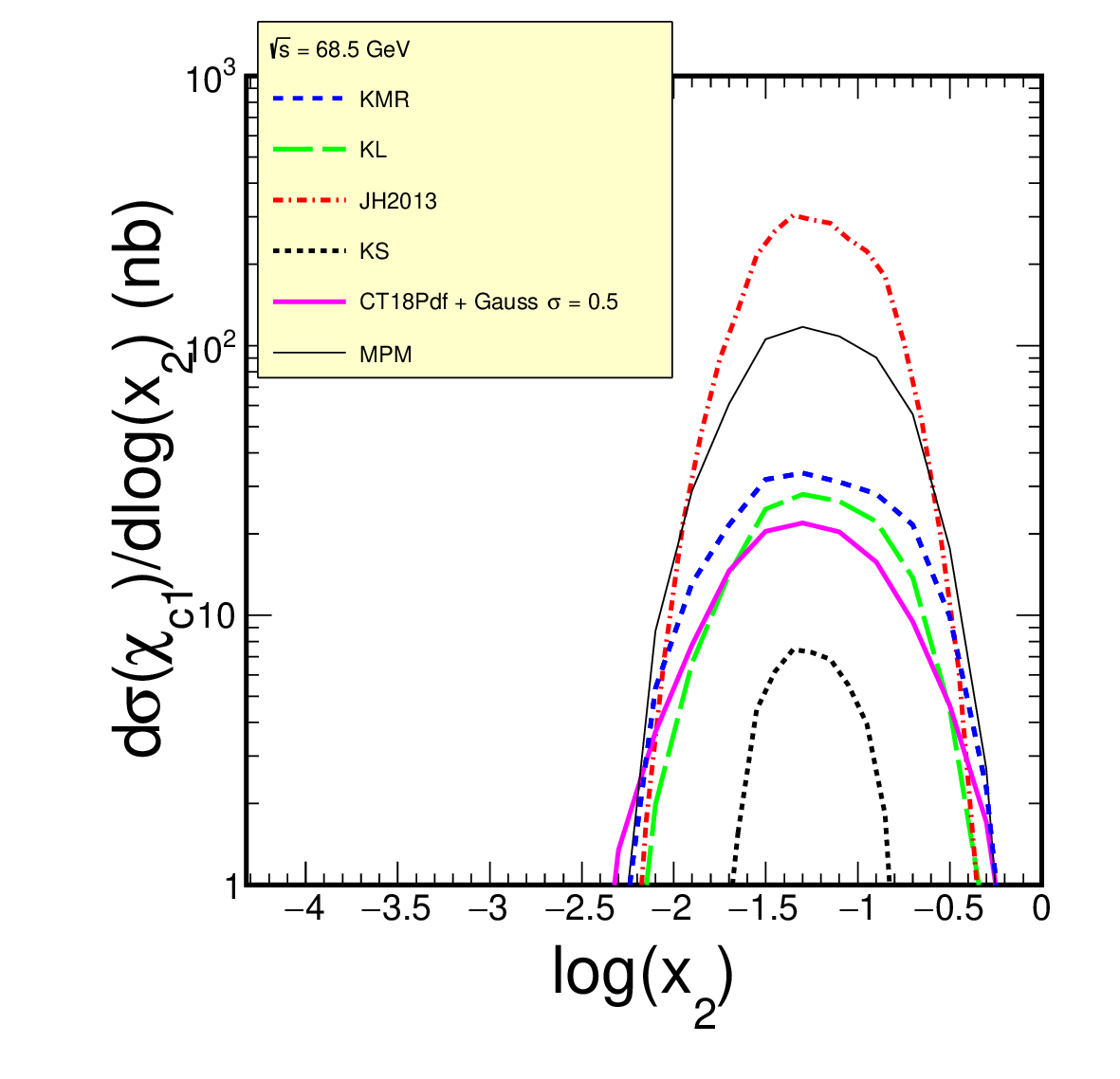}
\includegraphics[width=7.cm]{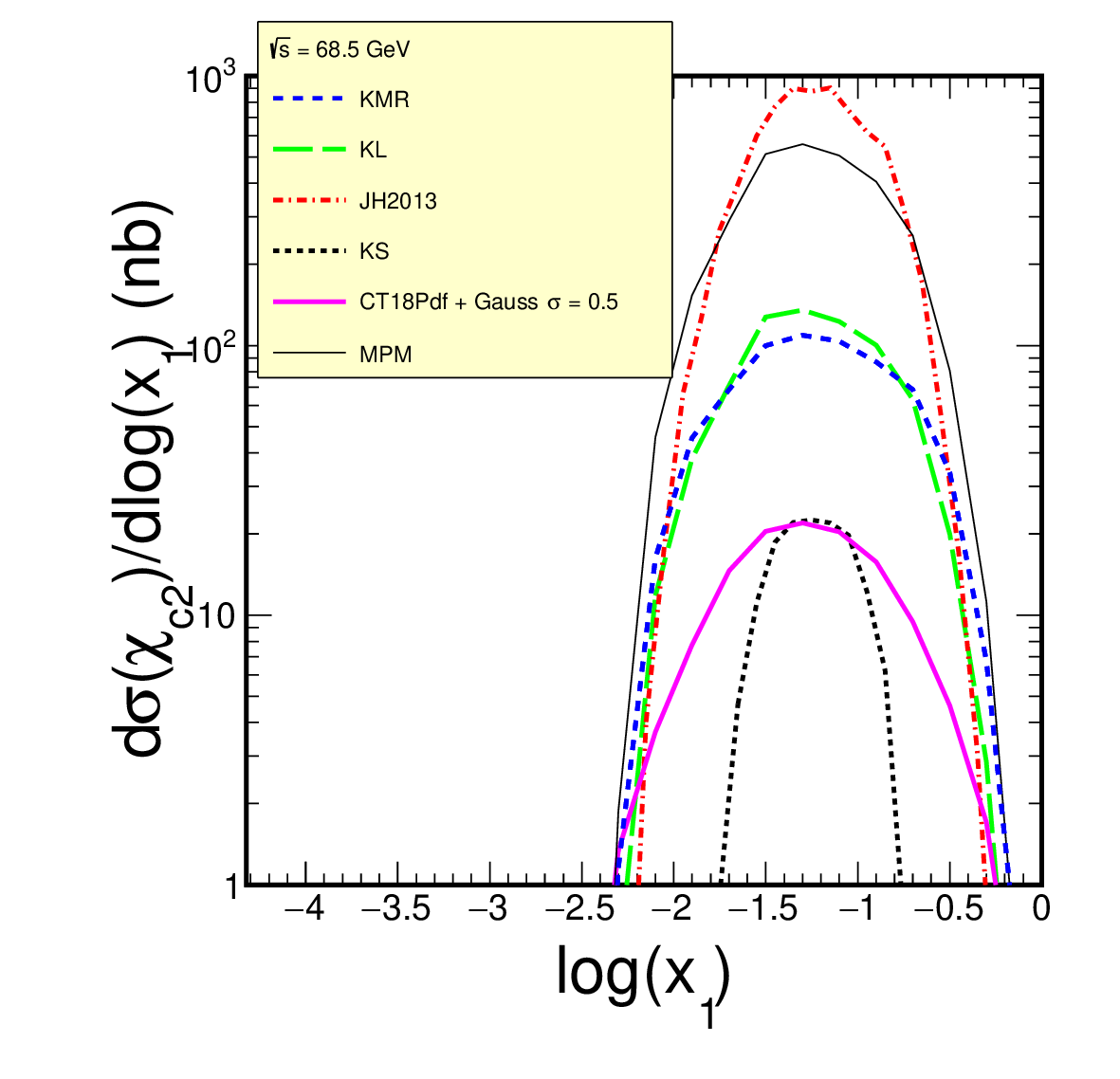}
\includegraphics[width=7.cm]{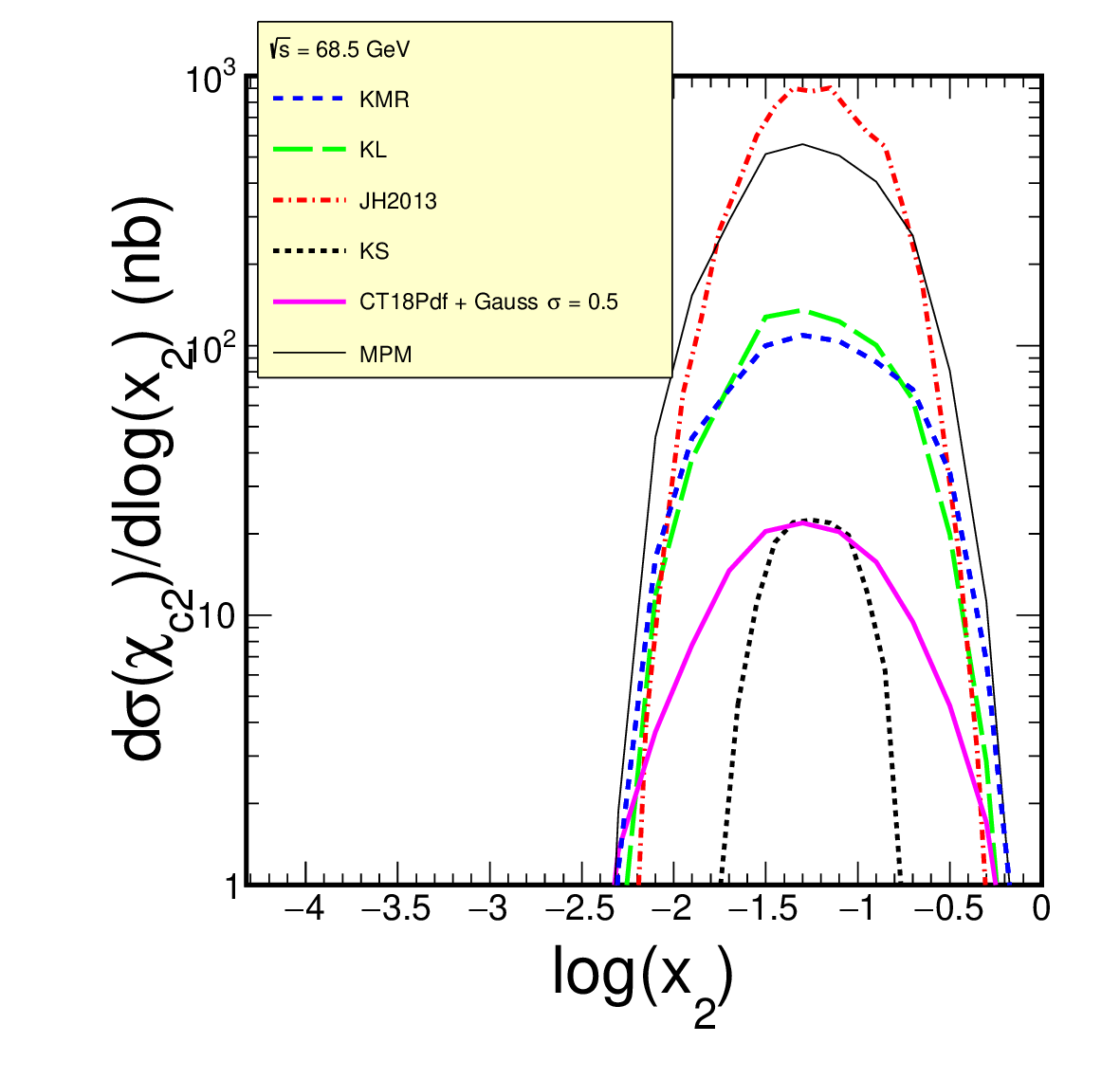}
\caption{$x_1$ (left) and $x_2$ (right) distributions for 
production of $\chi_c(1^+)$ (top) and $\chi_c(2^+)$ (bottom)
for different UGDFs.
In this calculation $\sqrt{s}$ = 68.5 GeV.}
\label{fig:x1x2_chic}
\end{figure}

In Fig.\ref{fig:x1x2_chic} we show distributions in 
$x_1$ or $x_2$ for production of $\chi_c(1^+)$ (top)
and $\chi_c(2^+)$ (bottom) mesons for the fixed target experiment
at $\sqrt{s}$ = 68.5 GeV. At $\sqrt{s}$ = 68.5 GeV one tests
UGDFs at $x_1, x_2 \sim$ 10$^{-2}$ - 10$^{-1}$.
We show distributions for different UGDFs. Some of them are
relevant for small-$x$, usually $x <$ 10$^{-2}$, some are relevant 
in a broader range of $x$.
Already here one sees huge span of the cross section, when using
the different UGDFs. Therefore the fixed target measurements at 
the LHC are very useful for testing UGDFs.

\begin{figure}[h]
\centering
\includegraphics[width=7.cm]{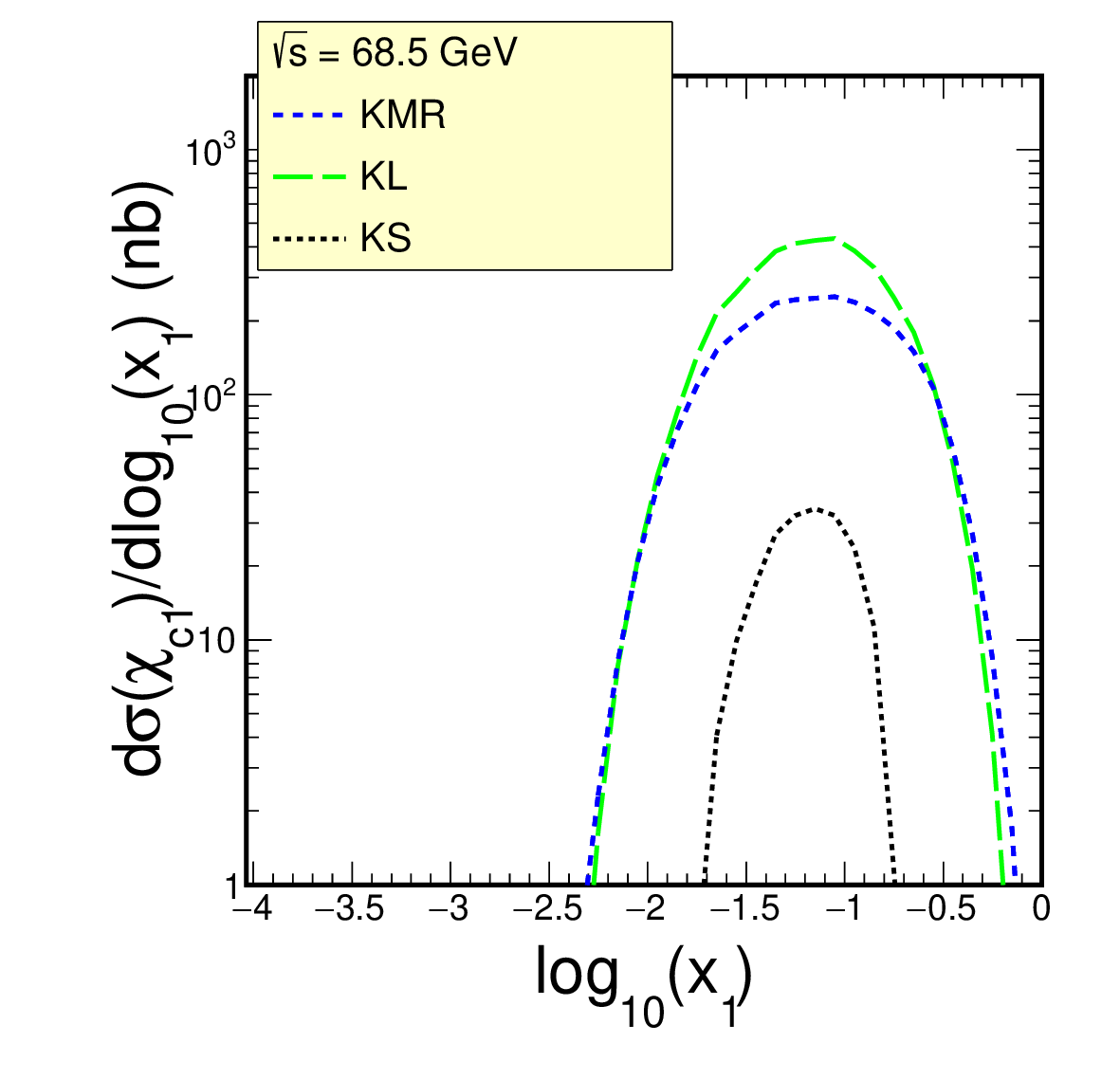}
\includegraphics[width=7.cm]{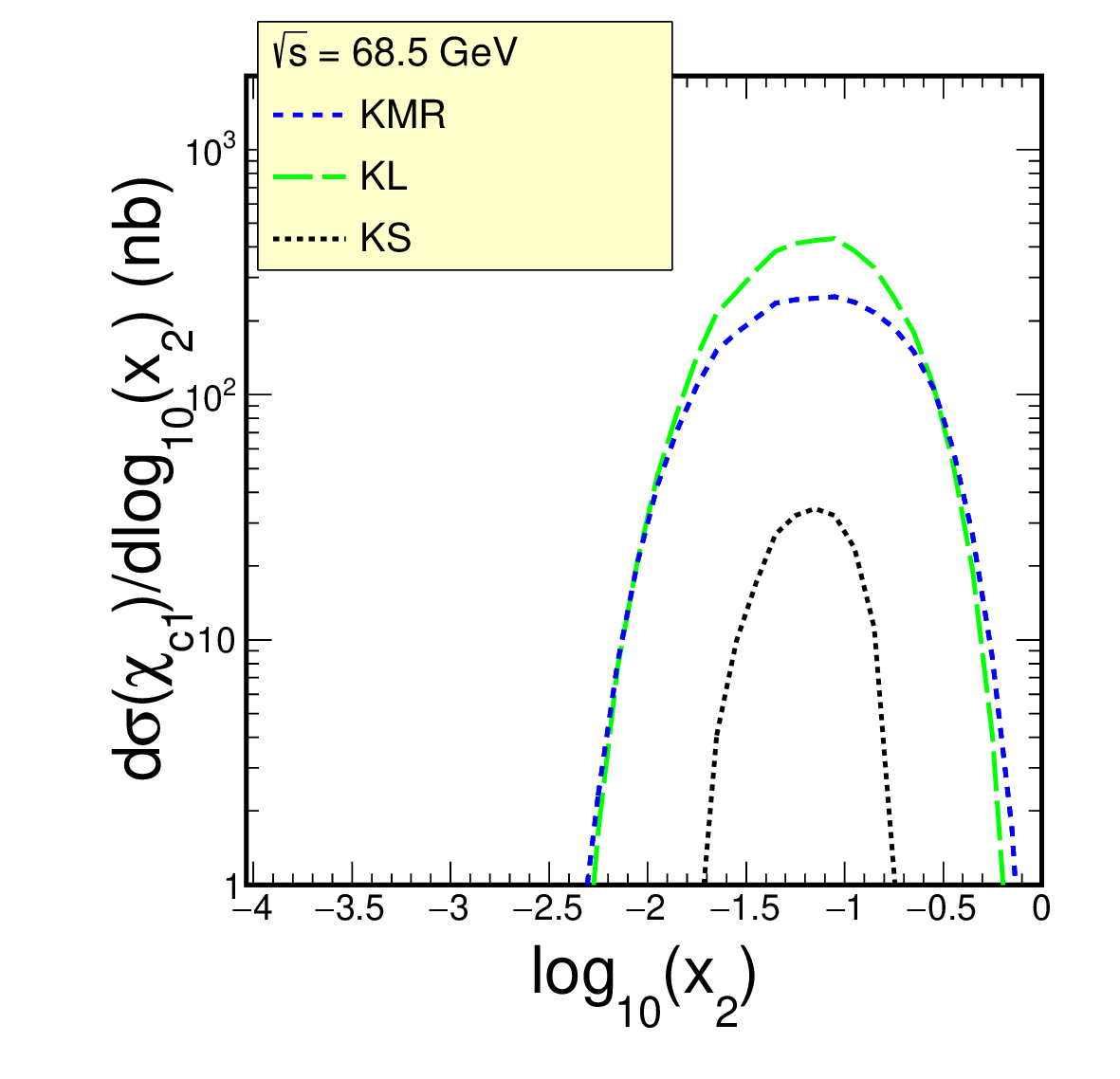}
\caption{$x_1$ (left) and $x_2$ (right) distributions
for direct production of $J/\psi$ for different UGDFs. 
Here $\sqrt{s}$ = 68.5 GeV.}
\label{fig:x1x2_jpsi}
\end{figure}

In Fig.\ref{fig:x1x2_jpsi} we show similar distributions for
direct production of $J/\psi$ (Fig.\ref{fig:gg_Jpsig}). 
The range is of course similar as for the $\chi_c$ production 
(compare Fig.\ref{fig:x1x2_jpsi} and Fig.\ref{fig:x1x2_chic}).
We observe similar span of the results.

\begin{figure}[h]
\centering
\includegraphics[width=7.cm]{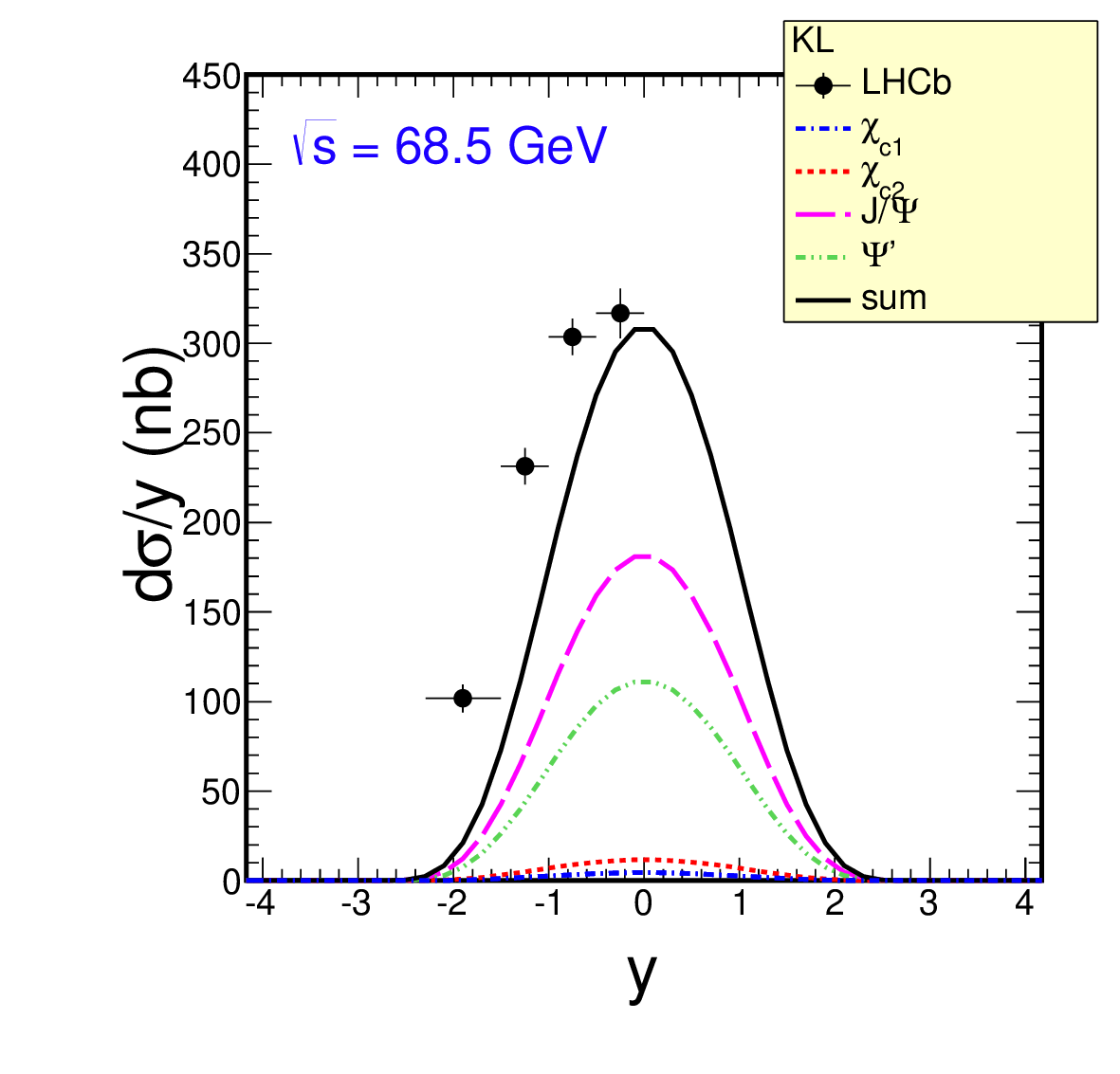}
\includegraphics[width=7.cm]{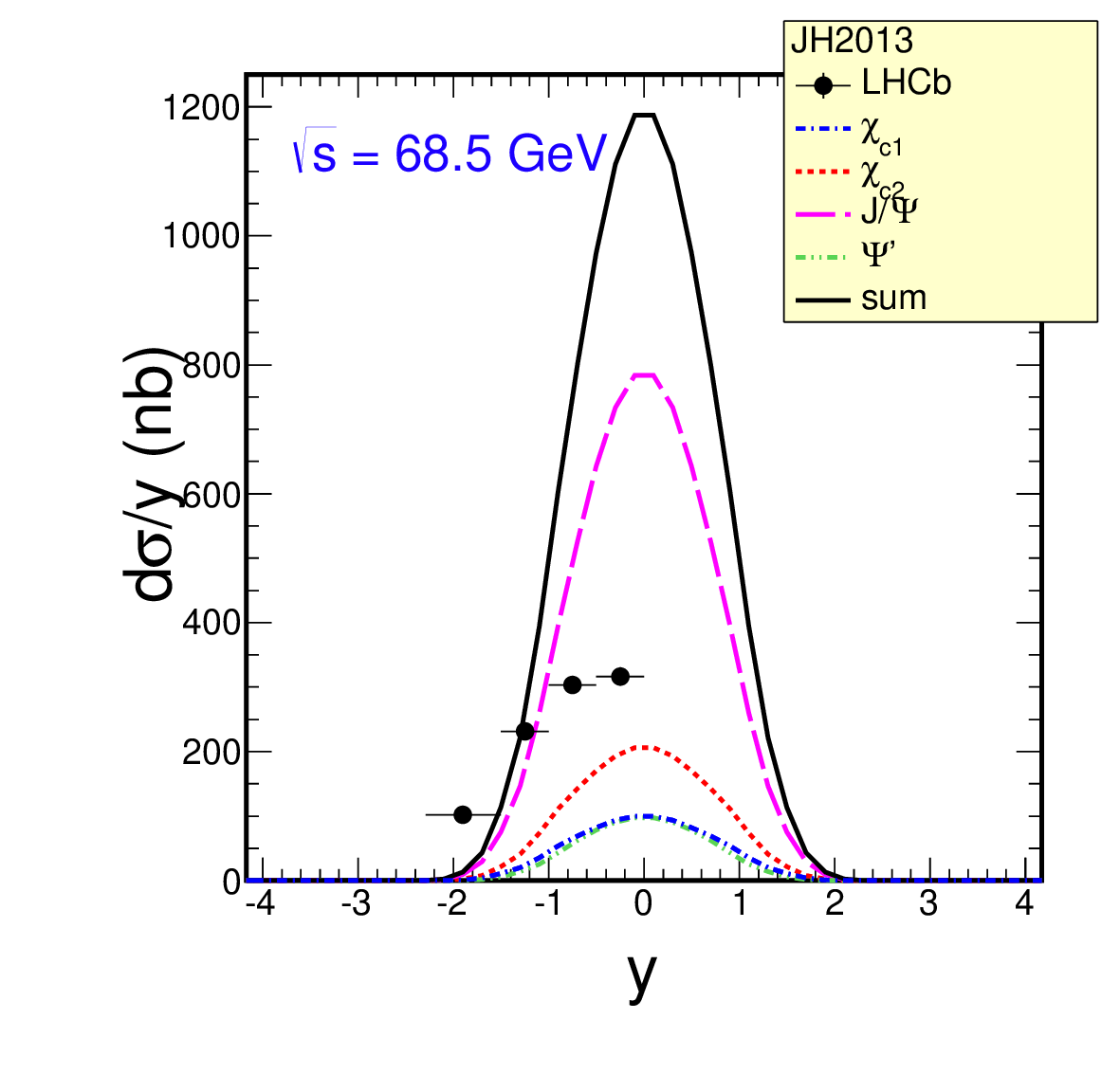}
\includegraphics[width=7.cm]{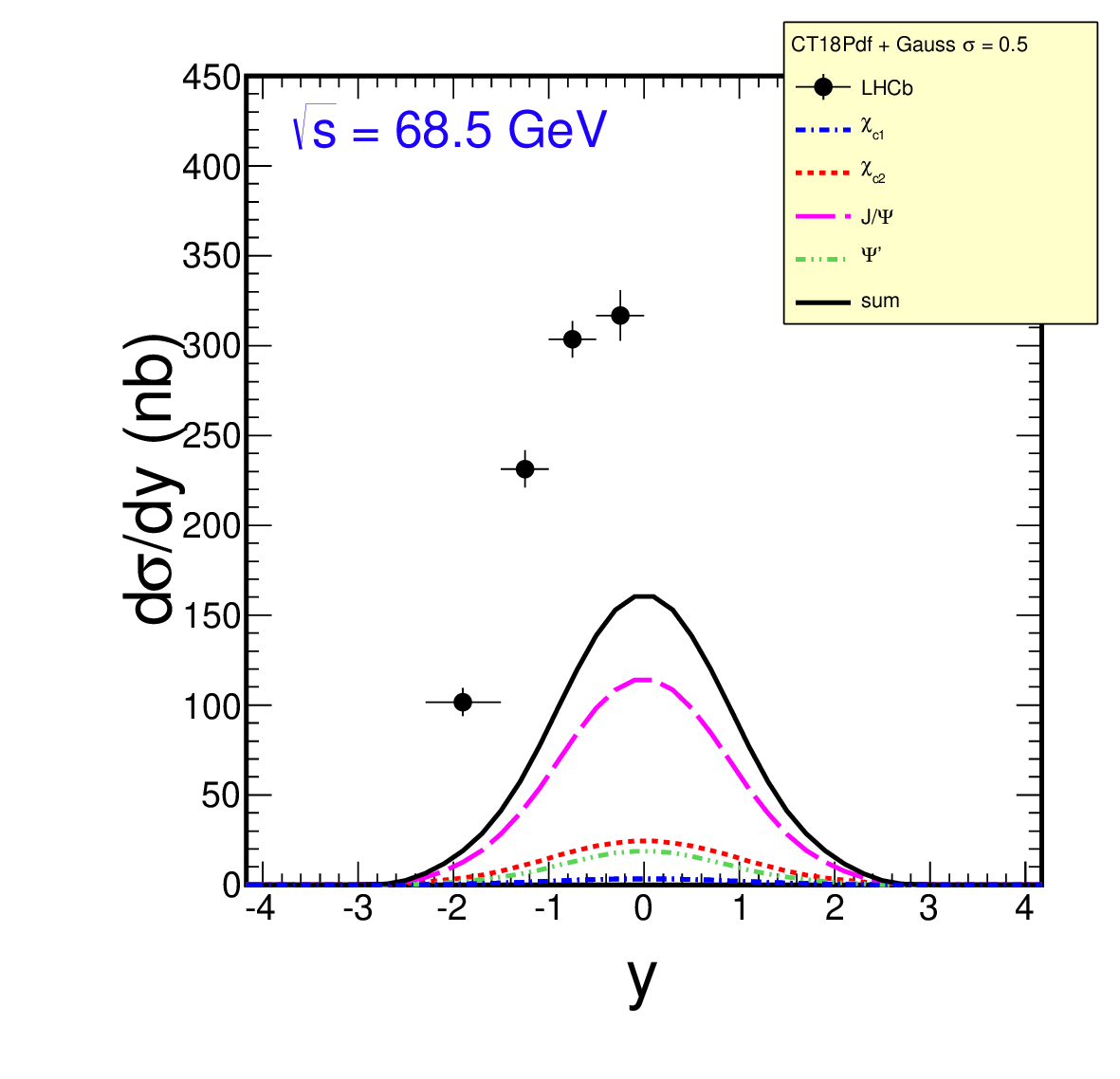}
\includegraphics[width=7.cm]{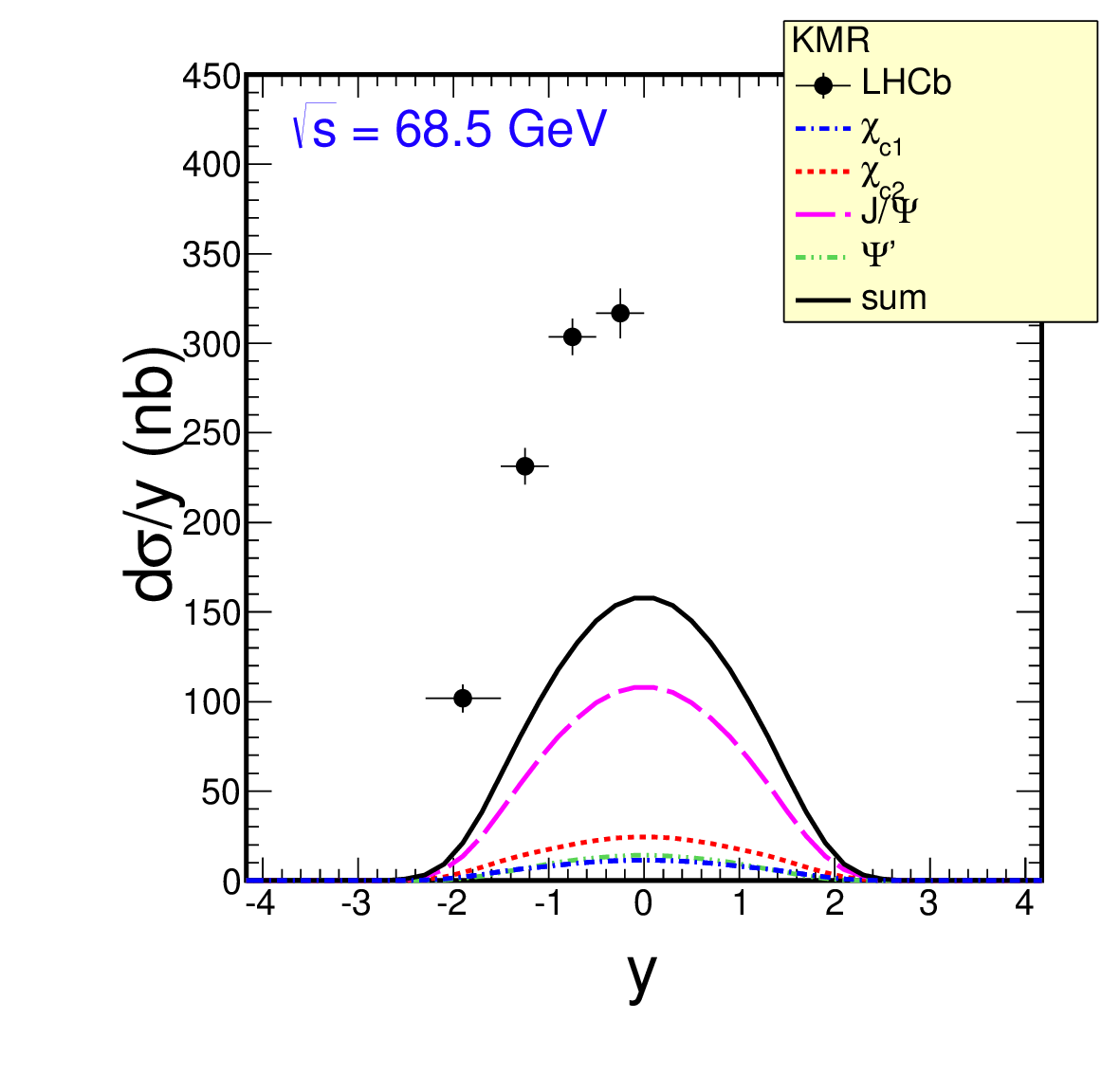}
\includegraphics[width=7.cm]{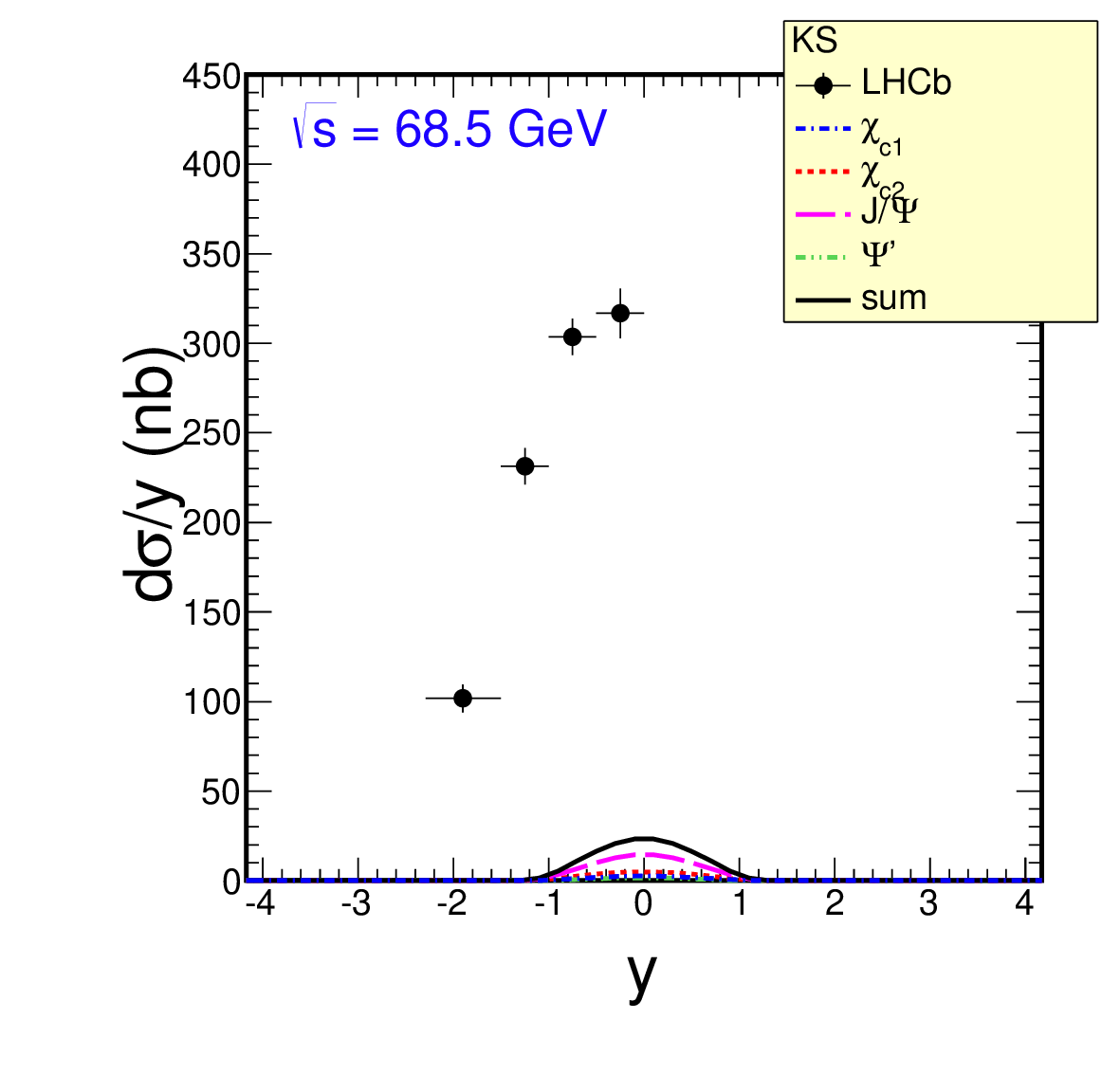}
\includegraphics[width=7.cm]{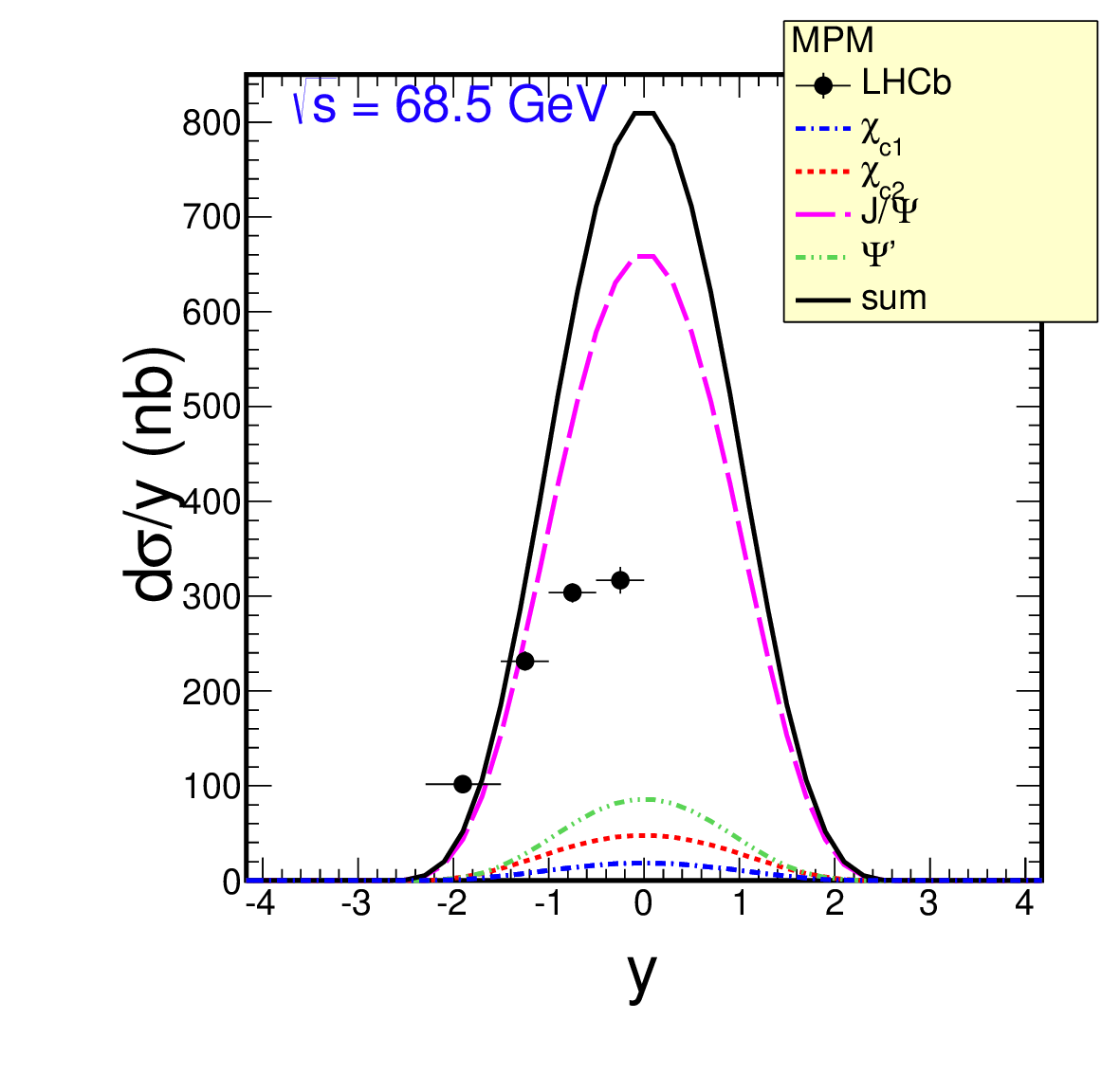}
\caption{Rapidity distribution of $J/\psi$ mesons for all considered mechanisms
for different unintegrated distribution functions.}
\label{fig_rapidity}
\end{figure}

Now we shall present distributions in measurable quantities.
In Fig.\ref{fig_rapidity} we present distribution in
rapidity of $J/\psi$ again for different UGDFs.

\begin{figure}[h]
\centering
\includegraphics[width=7.cm]{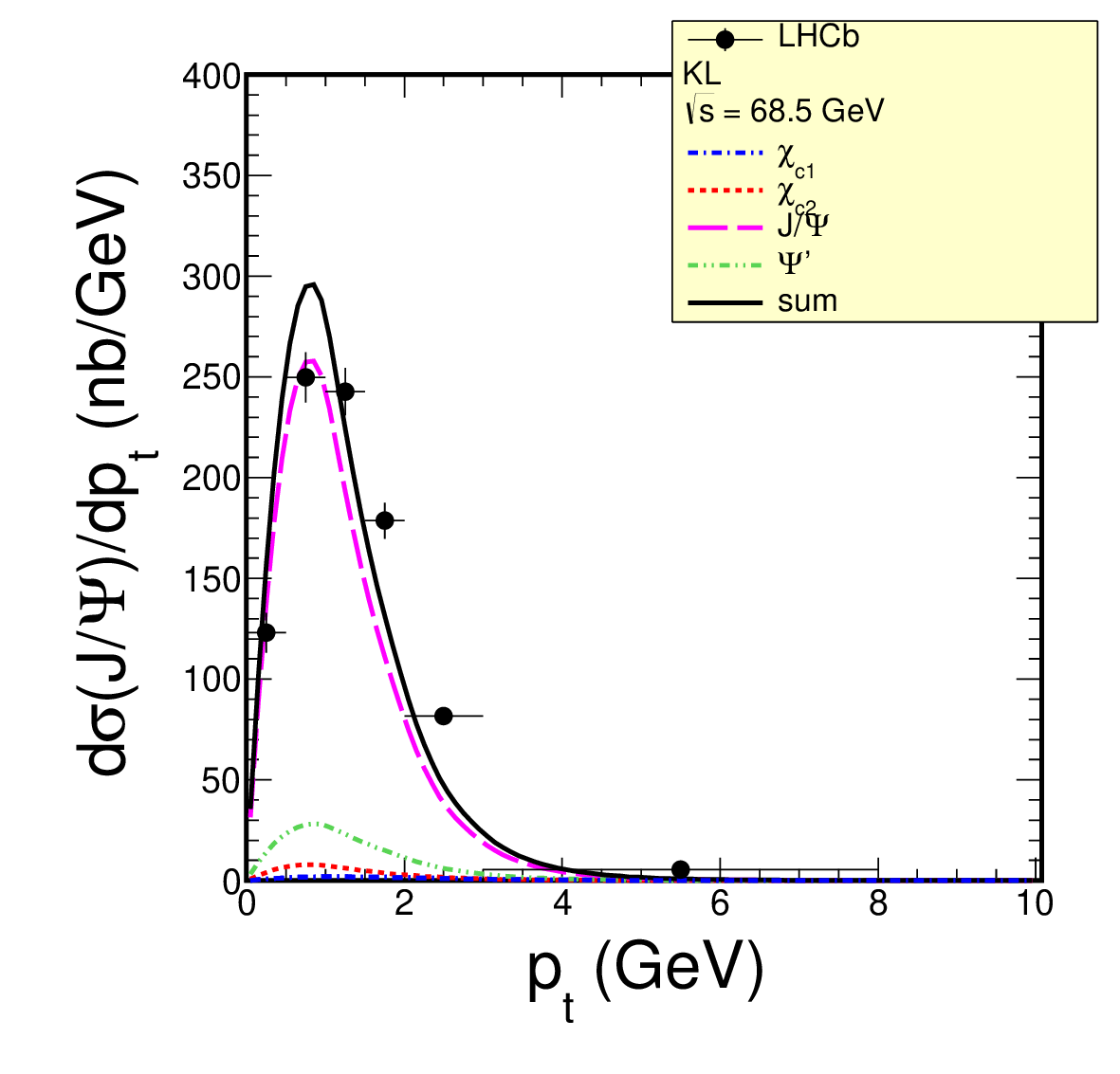}
\includegraphics[width=7.cm]{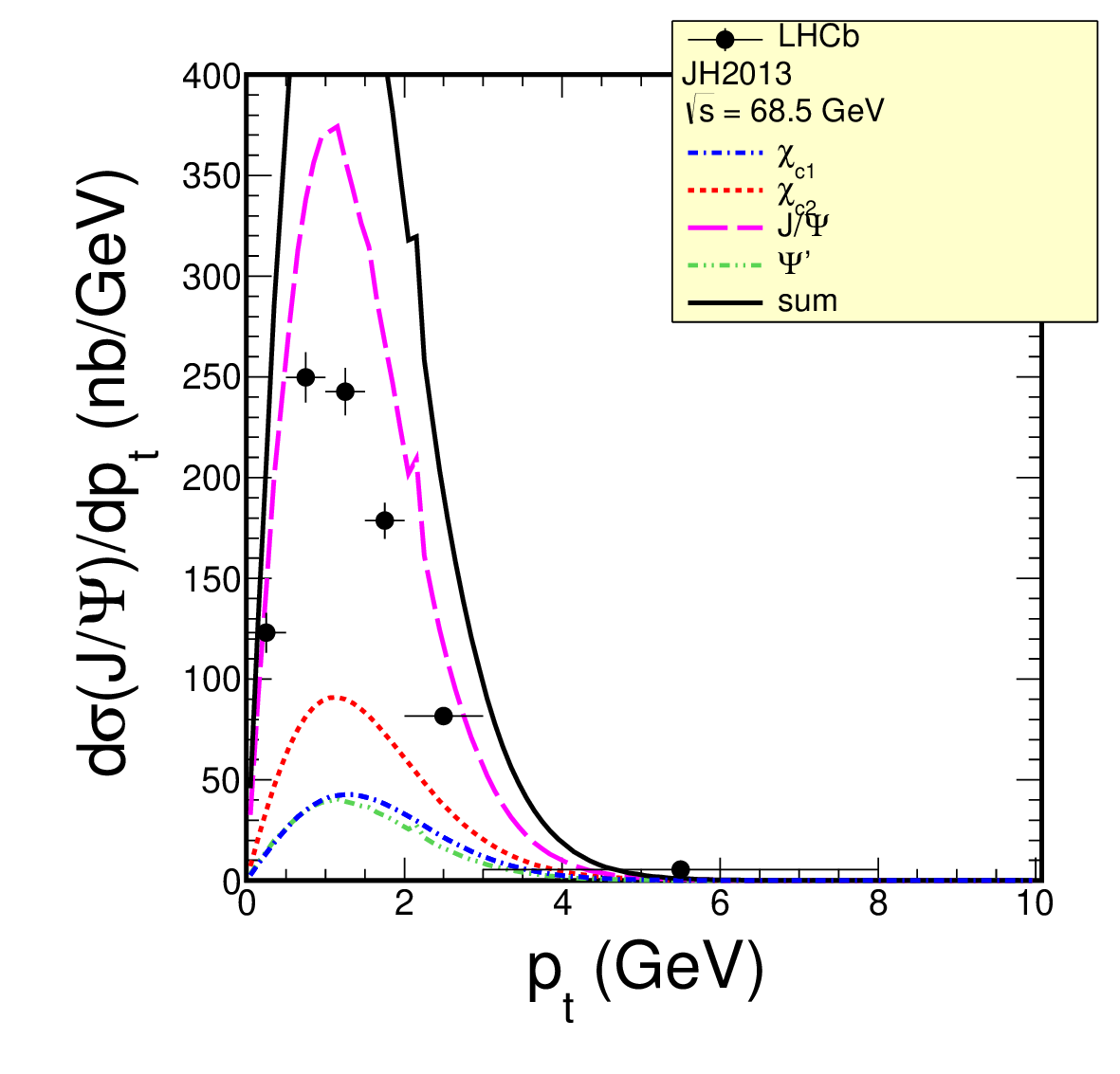}
\includegraphics[width=7.cm]{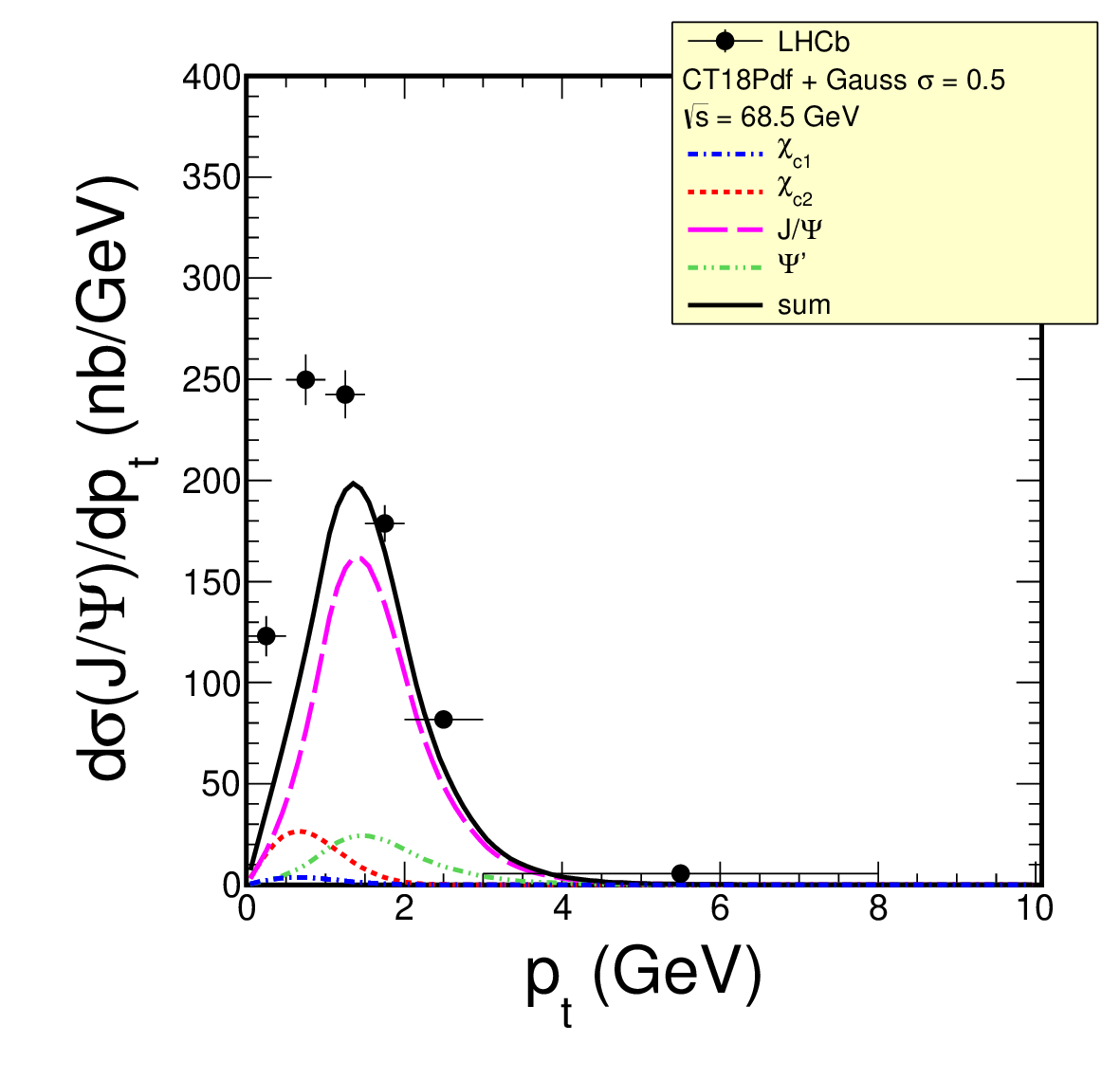}
\includegraphics[width=7.cm]{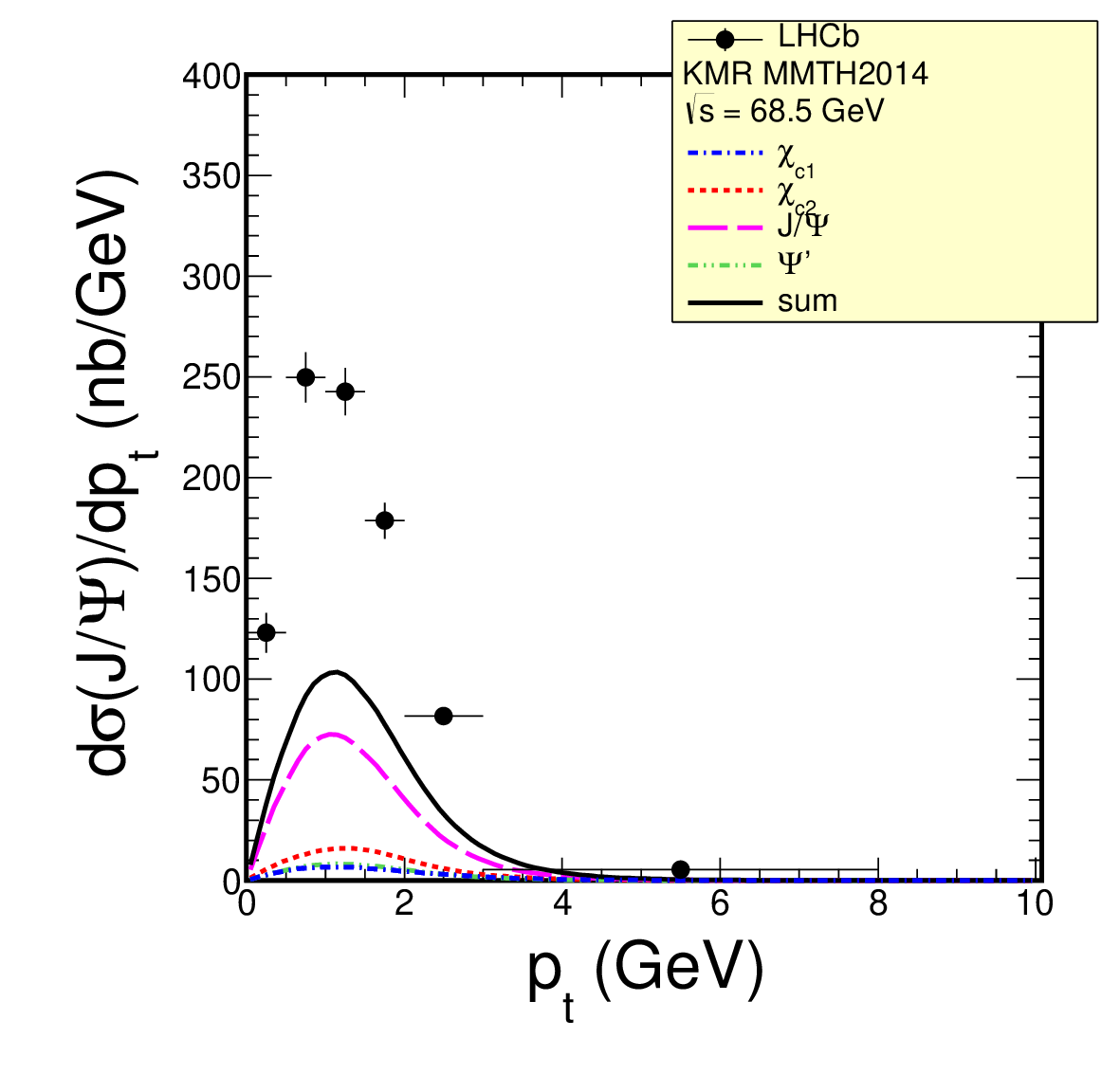}
\includegraphics[width=7.cm]{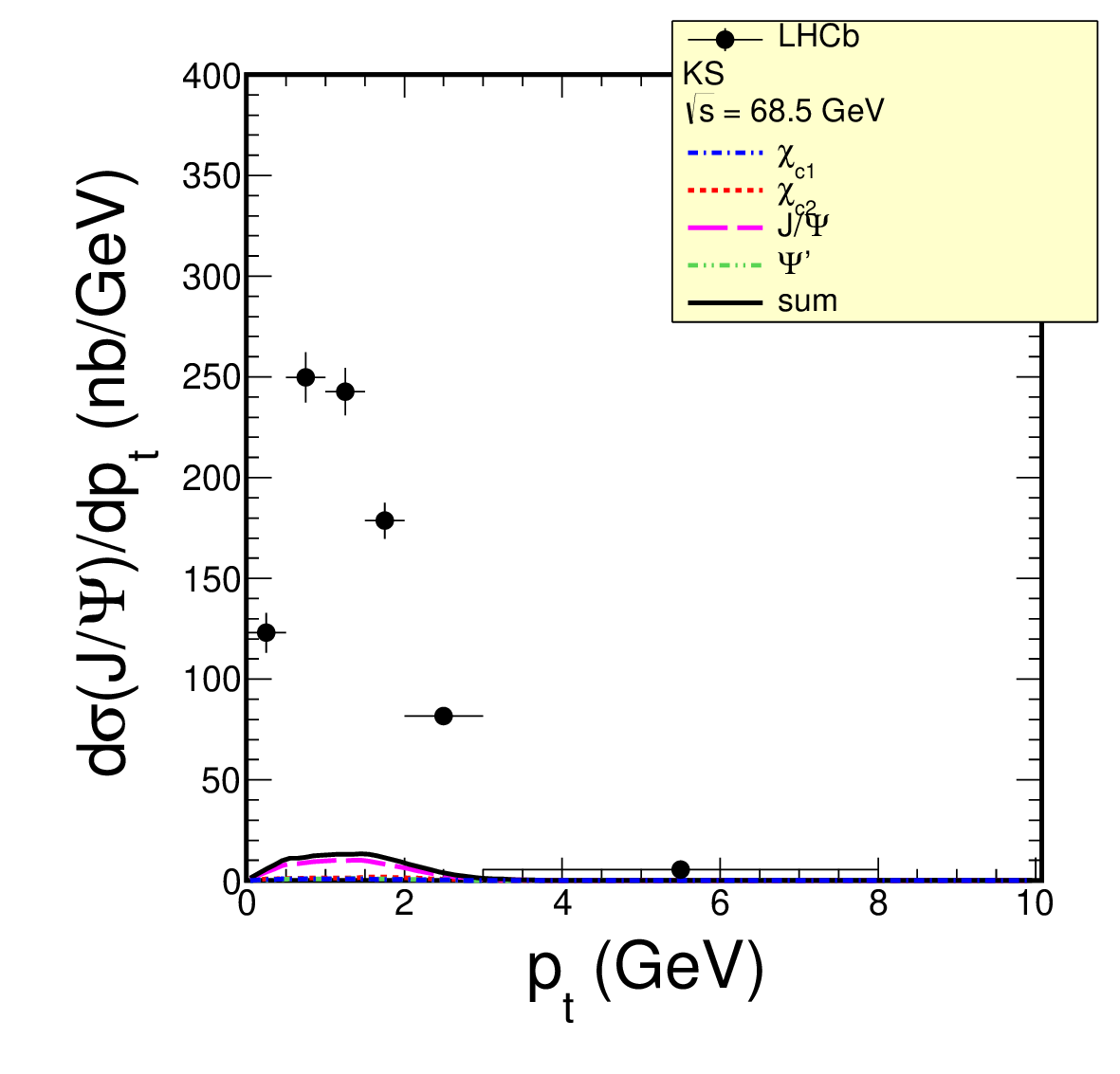}
\includegraphics[width=7.cm]{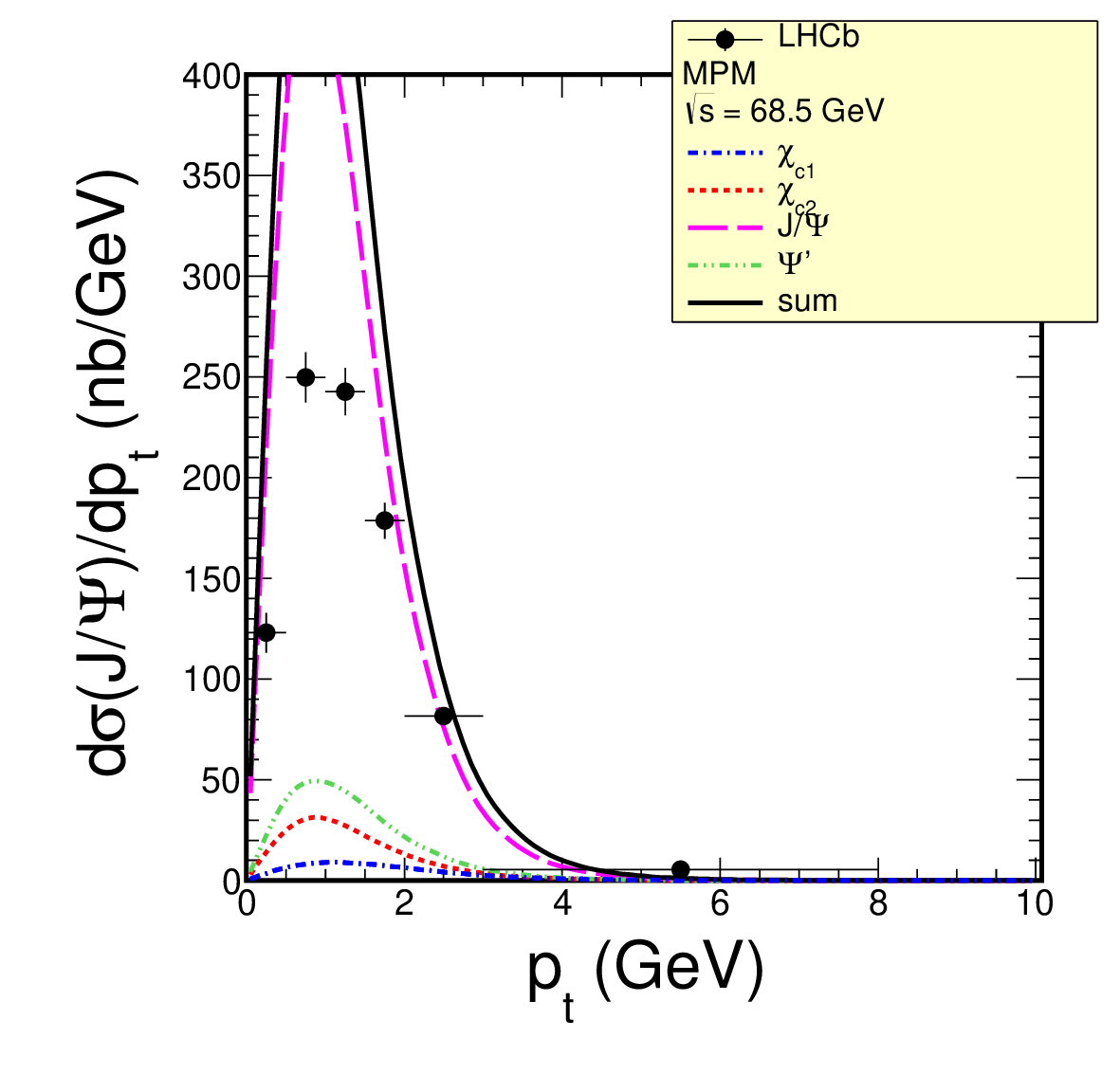}
\caption{Transverse momentum distribution of $J/\psi$ mesons for all
  considered mechanisms for different unintegrated distribution functions.}
\label{fig_pt}
\end{figure}

In Fig.\ref{fig_pt} we present our results 
for transverse momentum distributions of $J/\psi$ mesons.
In Fig.\ref{fig_rapidity} and Fig.\ref{fig_pt} we show separately 
both direct contributions and contributions from decays of 
$\chi_{c}(1)$, $\chi_{c}(2)$ and $\psi'$ mesons.
The solid lines represent the sum of these four components.
The numerical results are compared with experimental data of the
LHCb Collaboration.
For some UGDFs the cross section is too small, for others it is much 
too large, but for some of them the results are close to the 
experimental data.

\section{Conclusions}

In the present paper we have analyzed prompt production of $J/\psi$
quarkonia for energies corresponding to fixed target LHCb
p + A data. In that experiment relatively light nucleus
was used and the nuclear data was transformed to p+N collisions 
neglecting nuclear effects. The LHCb experiment corresponds to 
$\sqrt{s} =$ 68.5 GeV. 

Neglecting quark-induced processes that are still, for this low energy,
much smaller than the purely gluonic effects we have assumed the same 
mechanisms for $p+p$ and $p+n$ binary collisions. 
In this approximation one has therefore
$\sigma_{pA}^{J/\psi} \approx A \sigma_{pp}^{J/\psi}$, also 
for differential distributions.

In this exploratory studies we have therefore performed
calculations within $k_t$-factorization as we did previously 
in high-energy collisions \cite{CS2018}. The formalism used here
is therefore the same as there.
At high energies ($\sqrt{s} \sim 10$ TeV) one is sensitive to the region of very small
$x$ of the order of 10$^{-4}$. Here one is sensitive to much larger
values of $x_1$ or $x_2$, typicaly larger than 10$^{-2}$.
It is therefore interesting how the different UGDFs from 
the literature perform at the relatively low energies.

We have calculated distributions in rapidity and transverse momentum
of $J/\psi$. The obtained results have been compared to the LHCb data.
As one could expect there is relatively large spread of results
for this intermediate-$x$ region. The KMR and JH2013 uninterated
gluon distributions give the best description of the data. 
Quite good result has been obtained when
using extremely simple KL UGDF used previously to light particle
production.

\vskip 1cm

{\bf Acknowledgement}

We acknowledge a participation of Robert Brak in early stage
of this analysis.


\end{document}